\documentclass[usenatbib]{mnras}

\usepackage{natbib}
\usepackage{delarray}
\usepackage{color}
\usepackage{amsmath}
\usepackage{amssymb}
\usepackage{here}
\usepackage{graphicx}
\usepackage[utf8]{inputenc}
\usepackage{float}
\usepackage{epsfig}
\usepackage{tensor}
\usepackage{xspace}
\usepackage{multirow}

\usepackage[T1]{fontenc}
\usepackage{ae,aecompl}

\newcommand{\mach}{\mathcal{M}}
\newcommand{\xvect}{\mathbf{x}}

\newcommand{\be}{\begin{equation}} \newcommand{\ee}{\end{equation}}

\newcommand{\solarmass}{\mathrm{M}_{\sun}}
 
\newcommand{\pc}{\mathrm{pc}}
\newcommand{\kelvin}{\mathrm{K}}

\newcommand{\dderiv}{\mathrm{d}}

\newcommand{\acknowledgments}{\begin{small}\section*{Acknowledgments}\end{small}}
\newcommand\altaffilmark[1]{$^{#1}$}
\newcommand\altaffiltext[1]{$^{#1}$}

\newcommand{\framework}{GH15\xspace}
\newcommand{\Markfeedback}{K11\xspace}

\newcommand{\myquote}[1]{``#1''}

\title{The Necessity of Feedback Physics in Setting the Peak of the Initial Mass Function}

\author[Guszejnov, Krumholz \&\ Hopkins ]{
\parbox[t]{\textwidth}{ D\'avid Guszejnov\altaffilmark{1}\thanks{E-mail:guszejnov@caltech.edu}, Mark R. Krumholz\altaffilmark{2} and Philip F. Hopkins\altaffilmark{1}}
\vspace*{6pt} \\
\altaffiltext{1}{TAPIR, MC 350-17, California Institute of Technology, Pasadena, CA 91125, USA} \\
\altaffiltext{2}{Department of Astronomy \& Astrophysics, University of California, Santa Cruz, CA 95064 USA} 
}

\date{To be submitted to MNRAS, \today \vspace{-0.6cm}}
\begin{document}
\maketitle
\label{firstpage}
\hyphenation{Heating}

\begin{abstract}

A popular theory of star formation is gravito-turbulent fragmentation, in which self-gravitating structures are created by turbulence-driven density fluctuations. Simple theories of isothermal fragmentation successfully reproduce the core mass function (CMF) which has a very similar shape to the initial mass function (IMF) of stars. However, numerical simulations of isothermal turbulent fragmentation thus far have not succeeded in identifying a fragment mass scale that is independent of the simulation resolution. Moreover, the fluid equations for magnetized, self-gravitating, isothermal turbulence are scale-free, and do not predict any characteristic mass. In this paper we show that, although an isothermal self-gravitating flow does produce a CMF with a mass scale imposed by the initial conditions, this scale changes as the parent cloud evolves. In addition, the cores that form undergo further fragmentation and after sufficient time forget about their initial conditions, yielding a scale-free pure power-law distribution $\dderiv N/\dderiv M\propto M^{-2}$ for the stellar IMF. We show that this problem can be alleviated by introducing additional physics that provides a termination scale for the cascade. Our candidate for such physics is a simple model for stellar radiation feedback. Radiative heating, powered by accretion onto forming stars, arrests the fragmentation cascade and imposes a characteristic mass scale that is nearly independent of the time-evolution or initial conditions in the star-forming cloud, and that agrees well with the peak of the observed IMF. In contrast, models that introduce a stiff equation of state for denser clouds but that do not explicitly include the effects of feedback do not yield an invariant IMF.

\end{abstract}

\begin{keywords}
stars: formation -- turbulence -- galaxies: evolution -- galaxies: star formation -- cosmology: theory
\vspace{-1.0cm}
\end{keywords}

\section{Introduction}\label{sec:intro}

New stars form in dense molecular clouds as self-gravitating subregions collapse. Turbulent fragmentation is thought to be the main driving force of this process: turbulence compresses the gas, creating local density fluctuations that may be large enough to become self-gravitating. The appeal of this model comes from the fact that supersonic turbulence naturally produces a power-law relationship between velocity dispersion and size scale that is in good agreement with observations of molecular clouds (\citealt{Larson_law, Bolatto_2008, Kritsuk_larson_supersonic_origin}). A second advantage of a turbulence-based model is its universality. The initial mass function (IMF) of stars is observed to be close to universal (\citealt{IMF_review, IMF_universality}), with a high-mass end that is well described by a power-law with a slope of roughly $M^{-2.35}$ (\citealt{Salpeter_slope}) and a turnover at a few tenths of a Solar mass\footnote{Due to the high uncertainty of measurements of brown dwarfs the functional form of the turnover is not obvious from the data (\citealt{IMF_universality, SF_big_problems}). The most common fits are either a broken power law (\citealt{Kroupa_IMF}) or a lognormal (\citealt{Chabrier_IMF}).}. The mass at which this turnover occurs is robustly determined to be a few tenths of a Solar mass in all resolved stellar populations in the Milky Way (e.g., Figure 2 of \citealt{IMF_universality}) and in nearby galaxies (e.g., \citealt{Geha_turnover_univ}). Only a few resolved systems show even minor deviations in the location of the peak, and even then only by a factor of $\sim 2$ (e.g., $0.6-0.8\,\solarmass$ in Taurus -- \citealt{Luhman_Taurus}). This lack of variation is remarkable, given that the star-forming systems over which it is measured span many orders of magnitude in mass and density. Even in the most extreme environments, such as the cores of giant elliptical galaxies, the IMF turnover mass differs from the one found locally by at most a factor of a few (\citealt{Conroy_vanDokkum_ellipticals, Cappellari_IMF_var_2012, Geha_turnover_univ}). Such a universal distribution is most naturally explained by simple, universal physics that is independent of galactic environment, and the physics of turbulence is an obvious candidate.

There have been many attempts to formulate an analytic theory for the IMF, and for its turnover in particular, based on turbulence. Most are based on the random field approach first used in cosmology by \cite{PressSchechter}. This method was first applied to explain the IMF by \cite{Padoan_theory} and \cite{Padoan_Nordlund_2002_IMF}, then made more rigorous by \cite{HC08, HC2009, HC_2013} and \cite{excursion_set_ism}. Using such a model \cite{core_IMF} calculated the mass function of non-fragmented bound structures at a fixed time instant, a real life equivalent of which would be the core mass function (CMF), as opposed to the IMF. The observed CMFs of nearby star-forming regions have functional forms similar to that of the IMF (\citealt{Alves_CMF_IMF_obs, Rathborne_CMF, Sadavoy_observed_CMF}), with a Salpeter-like power-law at high masses and a turnover at lower masses, though the existence and the exact location of the CMF turnover are both quite uncertain due to issues of completeness and confusion – see \cite{IMF_universality} and \cite{SF_big_problems} for more discussion. The CMF derived by \cite{core_IMF} shares these features. The turnover mass in this model is set by the sonic mass $M_{\rm sonic}\sim c_s^2 R_{\rm sonic}/G$ that corresponds to the scale below which self-gravitating structures are subsonic; a similar mass scale arises in the Hennebelle \& Chabrier model, and one can show that the Hennebelle \& Chabrier and Hopkins mass scales are in fact identical up to constants of order unity (\citealt{SF_big_problems}). However, this scale is not universal, as it depends on the initial conditions in the star-forming cloud, which calls into question whether such a model can truly explain the near-universality of the IMF.

The proposition that the IMF is determined by the CMF, which in turn is set by the physics of isothermal gravito-turbulent fragmentation, has the appeal of simplicity. However, there remains an obvious question: once a core forms, why should one assume that it will collapse to form a single star, rather than fragmenting further? Simulations of isolated isothermal cores suggest exactly the latter (fragmented) outcome (e.g., \citealt{Goodwin04a, Dobbs06a, Walch12a}). In principle, the question of the fate of isothermal cores should be resolvable by simulations. In practice, however, this turns out to be a formidable technical challenge. Isothermal turbulence is scale-free (\citealt{McKee_ambipolar_scaling, SF_big_problems}), and thus it is not obvious what dynamic range is required to obtain a converged numerical result. To date, \emph{no} published simulation of isothermal gravito-turbulent fragmentation has demonstrated that the spectrum of point masses it produces is numerically converged, and those few authors who have attempted convergence studies (\citealt{Martel_numerical_sim_convergence, Kratter10a}) report non-convergence to the highest numerical resolutions probed. One possible explanation, advanced by \cite{SF_big_problems}, is that the characteristic structures created by isothermal turbulence are not singular points but singular filaments. Simulations produce filaments down to the smallest size scales they reach, and then the sink particle algorithm they use to represent collapsing regions breaks those filaments up into points at the grid scale.

Given these problems with purely isothermal fragmentation, a number of authors have proposed that the fragmentation cascade is arrested when the gas begins to heat up, in which case the characteristic stellar mass is determined by whatever physics causes the deviation from isothermality. The most common approach to this problem has been to adopt an equation of state that ``stiffens'' (i.e., the temperature begins to rise) above some characteristic density or surface density. Since super-isothermal gas is resistant to further fragmentation, one then identifies the IMF peak with the Jeans mass at this ``stiffening density'', on the basis that fragmentation will be suppressed beyond that point (\citealt{Whitworth98a, Larson2005}).

Stiffening of the equation of state can be caused by a diverse range of processes, including the inability of radiative cooling to keep up with adiabatic heating at a density $n\sim 10^{10}$ cm$^{-3}$ (e.g., \citealt{Masunaga1998}) or a surface density $\Sigma\sim 5000$ $\solarmass\,\pc^{-2}$ (e.g., \citealt{Glover_EQS_lowgamma_ref}), the onset of dust-gas coupling at a density $n\sim 10^5$ cm$^{-3}$ (\citealt{Larson2005, Elmegreen08a}), or combinations of the above (e.g., \citealt{Spaans00a}). Numerical simulations based on these equations of state do find a converged mass scale that can plausibly be identified as a characteristic mass for the IMF (e.g., \citealt{Bonnell06a, Bate09a}). However, it is not clear that the mass scale introduced by those models is actually universal (as opposed to set by initial conditions). Moreover, a number of authors have pointed out that radiative feedback is likely to be more important than any of these processes in setting the gas temperature in an actively star-forming region, and that this process is not well-described by an equation of state (\citealt{Krumholz06a, Krumholz07a, Offner09a, Urban10a}). Simulations of star cluster formation including radiative feedback suggest that it is capable of producing an IMF peak that is numerically converged and relatively insensitive to changes in interstellar conditions (\citealt{Bate_2009_rad_importance, Bate12a, Bate2014, Myers11a, Krumholz11a, Krumholz12a}).

While these developments are promising, the numerical expense of large-scale simulations including radiative feedback means that only a very small number of calculations have been performed. Moreover, analytic models of fragmentation with radiative feedback have, up to this point, been quite simple (e.g., \citealt{Krumholz08a, Krumholz11a}), and have not been linked to an analytic theory for the full IMF. Recently, \cite{TurbFramework} introduced a new method for performing semi-analytic calculations of turbulent fragmentation. Crucially, this method retains spatial information about how gas fragments, making it possible to include localized feedback mechanisms like stellar radiative heating. These calculations are rapid, enabling a much broader exploration of parameter space than can yet be accomplished with full three-dimensional radiation-hydrodynamic simulations.

In this paper we combine the \cite{TurbFramework} fragmentation model with the \cite{Krumholz_stellar_mass_origin} model for stellar radiative feedback (henceforth referred to as \framework and \Markfeedback respectively). We also explore alternative treatments of gas thermodynamics, including both isothermal and stiff equation of state models. We use these method to study the predicted IMF in a wide variey of star-forming environments. The remainder of this paper is laid out as follows. First, in Sec. \ref{sec:method} we introduce the semi-analytical framework we use to test different models of star formation. In Sec. \ref{sec:isothermal} we show that isothermal turbulent fragmentation leads to a scale-free IMF. In Sec. \ref{sec:EOS} we show that models with a stiffened equation of state are inherently sensitive to the initial conditions so they cannot provide an invariant mass scale. In Sec. \ref{sec:heating} we provide a simple model for protostellar heating that leads to an IMF with remarkably little sensitivity to initial conditions. Finally, in Section \ref{sec:conclusions} we discuss the implications of our findings, and conclude.

\section{Model and Methodology}\label{sec:method}

\subsection{Model Overview}

In order to test the different models we are using the semi-analytical framework of \framework. This takes advantage of the fact that the density fluctuations in a highly turbulent medium locally obey approximately lognormal random field statistics, thereby avoiding the need for computationally expensive hydrodynamical simulations while still preserving spatial information (unlike analytical excursion set models like \citealt{core_IMF, GuszejnovIMF}). The present version of the model only includes the bare essential physics: turbulence (pumped by the collapse of the cloud), collapse (at constant virial parameter, motivated by \citealt{Brant_turb_pumping} and \citealt{Murray_2015_turb_sim}), an equation of state (EOS) and a simple feedback prescription.

The initial conditions of clouds are defined by their mass, the sonic length ($R_{\rm sonic}$, scale at which the turbulent velocity dispersion is equal to the sound speed) and the sonic mass ($M_{\rm sonic}$), from which other parameters (e.g. temperature, Mach number) can be derived. For details about initial conditions see \framework (a detailed step-by-step guide to the model is provided in Appendix A).

Our simulations start from a giant molecular cloud (GMC) with fully developed turbulence and follow its collapse. Every time a new self-gravitating substructure appears (i.e., the cloud fragments) the code is run recursively for each substructure. When a cloud reaches the protostellar size scale ($\sim 10^{-4}\pc$) it is considered to be fully collapsed into a protostar and the simulation stops. This means that the final output of the code is the protostellar system mass function (PSMF) which we will assume to be identical to the IMF throughout this paper. This assumption is not quite accurate, particularly in the brown dwarf regime, as it neglects the production of brown dwarfs via disk fragmentation. We also neglect the growth of stars by Bondi-Hoyle accretion of gas that was not initially part of their collapsing core, though this effect is likely negligible (\citealt{krumholz05a}). The leftover unbound material is assumed to escape. All the simulations we present here start from $10^4\,\solarmass$ GMCs, as the results are completely insensitive to the size of the parent cloud (for demonstration see \framework).

\subsection{Equation of State Models}

In this paper we consider a series of models that include increasingly sophisticated treatments of gas thermodynamics. The simplest, which correspond to the usual assumption in turbulent fragmentation models, is that the gas is isothermal, corresponding to an adiabatic index $\gamma = 1$. The next level of complexity is simulations with a non-constant $\gamma$. The simulation allows for arbitrary equations of states which are taken into account as effective polytropes:
\begin{equation}
\label{eq:T_evol}
T(\xvect,t+\Delta t)=T(\xvect,t)\left(\frac{\rho(\xvect,t+\Delta t)}{\rho(\xvect,t)}\right)^{\gamma\left(t\right)-1},
\end{equation}
where $\gamma\left(t\right)$ is the effective polytropic index at the time $t$. To explore models in which the key physical process is a stiffening of the equation of state, we consider two possible formulations. Some authors have proposed that stiffening occurs at a characteristic surface density $\Sigma_{\rm crit}$, and we refer to models of this form as $\gamma(\Sigma)$ equations of state (EOS's). The particular parameterization we explore in this work is similar to that that proposed by \cite{Glover_EQS_lowgamma_ref}, which is 
\begin{equation}
\label{eq:gamma}
\gamma(\Sigma) =\begin{cases} 1.0 &\Sigma < \Sigma_{\rm crit} \\
															31/24 &\Sigma > \Sigma_{\rm crit}\\
		  \end{cases},
\end{equation}
where $\Sigma = M/(4 \pi R^2)$ for a cloud of mass $M$ and radius $R$\footnote{Note that the value of $31/24$ was chosen to allow the comparison of models with protostellar heating and $\gamma(\Sigma)$ EOSs (see Sec. \ref{sec:heating}). The choice of this value has no effect on the sensitivity of the results to initial conditions.}. \framework shows that using the standard value of $\Sigma_{\rm crit} = 5000\,\solarmass/\pc^2$ leads to a turnover mass of $\sim 0.01\,\solarmass$, much too low compared to the observed IMF; indeed, the mass picked out by this choice is simply the opacity limit for fragmentation (\citealt{Rees1976}). For this reason,  we set $\Sigma_{\rm crit}$ so that it is equal to critical surface density of the protostellar heating model ($\Sigma_{\rm heat}$, see Sec. \ref{sec:heating}) in the standard ($T_0=10\,\kelvin$) scenario. This means $\Sigma_{\rm crit}\sim 130\,\solarmass\pc^{-2}$. Using a higher surface density would only shift the turnover mass scale to lower values, it would not affect its sensitivity to initial conditions (see \framework for results with such an EOS). In other words, we are giving these models their ``best chance'' to fit the data.

Another formulation we consider is one where the stiffening occurs at a characteristic volume density $\rho_{\rm{crit}}$, which we refer to as a $\gamma(\rho)$ EOS. The form we adopt for this EOS is equivalent to the one used by \cite{Bate09a}:
\begin{equation}
\label{eq:gamma_rho}
\gamma(\rho) =\begin{cases} 1.0 &\rho < \rho_{\rm crit} \\
															1.4 &\rho > \rho_{\rm crit}\\
		  \end{cases},
\end{equation}
where $\rho = 3 M / 4\pi R^3$. Once again we chose the critical value so that it is convenient to compare with the other models so we set $\rho_{\rm crit}=15000\,\solarmass/\pc^{-3}$ corresponding to $n_{H_2,\rm crit}\approx 2.6\times 10^{5}\,\mathrm{cm}^{-3}$. 

For reference we also include a scenario with a more physically-motivated EOS based on the works of \cite{Masunaga_EQS_highgamma_ref} and \citealt{Glover_EQS_lowgamma_ref}:
\begin{equation}
\label{eq:gamma_phys_rho}
\gamma_{\rm phys}(\rho) =\begin{cases} 
0.8 &\rho < \rho_{\rm crit,1} \\
1.0 &\rho_{\rm crit,1}<\rho < \rho_{\rm crit,2} \\
1.4 &\rho > \rho_{\rm crit,2}\\
		  \end{cases},
\end{equation}
where we set $\rho_{\rm crit,1 }=5000\,\solarmass/\pc^{-3}$ and $\rho_{\rm crit,2 }=5\times 10^8\,\solarmass/\pc^{-3}$ corresponding to $n_{H_2,\rm crit,1 }\approx 10^{5}\,\mathrm{cm}^{-3}$ and $n_{H_2,\rm crit,2}\approx 10^{10}\,\mathrm{cm}^{-3}$. 

While these are only three of the EOS's that have been proposed in the literature, they serve as representative examples of the outcomes produced by such an approach.

\subsection{Radiation Feedback Models}

The final class of models we consider are those with a simple treatment of protostellar radiative feedback. In these we assume that the center of self-gravitating clouds collapses first, forming a protostellar seed, then the rest of the cloud collapses onto it
. The energy of the matter accreted by this seed is radiated within the optically thick core. The temperature of the material depends on the accretion rate onto the protostar (and thus the mass and dynamical time of the gas around it), and on the energy yield per unit mass from accretion, which we denote $\Psi$. The value of $\Psi$ is set by the protostellar mass-radius relation, and \Markfeedback shows that it is determined primarily by the effects of deuterium burning, which thermostats the central temperatures of protostars. Because deuterium burning begins when protostars are only a few $\times 10^{-2}\,\solarmass$, and, for low mass protostars, continues for $\sim 10$ Myr, it is the dominant factor in setting $\Psi$ during the bulk of a molecular cloud's star-forming history. Comparing with detailed protostellar evolution calculations, \Markfeedback finds that $\Psi\approx 2.5\times 10^{10}\,\rm{J/kg}$ to better than half a dex accuracy for all protostellar masses in the range $0.05-1\,\solarmass$, and to better than a dex accuracy from $0.01-0.05\,\solarmass$. We therefore adopt this value of $\Psi$ throughout the remainder of this paper. Following \Markfeedback, this heats any core harboring an accreting protostar up to a temperature
\begin{equation}
T_{\rm heat}^4\approx\frac{\Psi\sqrt{G}}{4\pi\sigma_{SB}}M^{3/2}R^{-7/2}.
\label{eq:T_heat}
\end{equation}
Crudely, this scaling reflects energy conservation as $L=4\pi R^2\sigma_{SB}T_{\rm heat}^4$ for the opaque cloud. Combined, internal heating and the physical processes captured by the EOS models set the temperature as
\begin{equation}
\label{eq:T_summing}
T^4=T_{\rm EOS}^4+T_{\rm heat}^4,
\end{equation}
 where $T_{\rm EOS}$ is the temperature of the cloud if only EOS effects are taken into account. Note that this is an extremely simplistic treatment of protostellar heating, where each cloud is assumed to have a protostar \myquote{seed} at its center which heats (uniformly) only its own cloud. This heating is assumed to be \myquote{turned on} as soon as the cloud forms. However, since the temperature depends on $\Psi$ only to the $1/4$ power, even a factor of $\sim 10$ error in its value, as can happen for $0.01\,\solarmass$ protostars, corresponds to a relatively modest error in $T$. For stars near the peak of the IMF, which we shall see this model places at $\sim 0.3$ $M_\odot$, the error is even smaller. 

To easily identify the results for different models and initial conditions we use the labels shown in Table \ref{tab:simparam}. The \textit{T10} label refers to initial conditions similar to MW GMCs,  \textit{T20} and \textit{T75} have enhanced temperatures as are typically found in regions of very active star formation or in the Galactic center, \textit{hiDens} has enhanced temperature and density, similar to a dense, massive star-forming region in the MW, while \textit{ULIRG} runs have the very high temperature and strong turbulence characteristic to the clouds of Ultra Luminous Infrared Galaxies (ULIRGs). Finally, the \textit{hiMach} model has an enhanced Mach number but fixed temperature; we are unaware of a physical analog for this case, but we include it because it provides useful insight into the physics of the model.

\begin{table*}
	\centering
		\begin{tabular}{ | c | c | c| c |c |c |c |c |c |}
		\hline
		\multirow{2}{*}{\bf Label} & \multicolumn{2}{c|}{\bf Input Parameters}  & \multicolumn{5}{c|}{\bf Derived Parameters}  &  \multirow{2}{*}{\bf Thermodynamics}  \\
		\cline{2-8}
		 & \bf $\mathbf M_{\rm sonic}$ [$\solarmass$]& \bf $\mathbf R_{\rm sonic}$ [pc] & \bf $\mathbf T_0$ [K] & \bf $\mathbf R_0$ [$\pc$] & \bf $\mathbf \mach_0$ & \bf $\mathbf \Sigma_0$ [$\solarmass\pc^{-2}$] & \bf $\mathbf n_{\rm{H}_2,0}$ [$\mathrm{cm}^{-3}$] &   \\
		\hline
		\hline
		IsoTherm\_T10 & 2.3 & 0.1 & 10 & 9.3 & 9.6 & 9.3 & $50$ & Isothermal  \\
		\hline
		\hline
		EOS$\mathrm{\Sigma}$\_T10 & 2.3 & 0.1 & 10 & 9.3 & 9.6 & 9.3 & $50$ & $\gamma(\Sigma)$ EOS  \\
		\hline
		EOS$\mathrm{\Sigma}$\_T20 & 4.6 & 0.1 & 20 & 6.5 & 8.1 & 18.6 & $150$ & $\gamma(\Sigma)$ EOS  \\
		\hline
		EOS$\mathrm{\Sigma}$\_T75 & 17.25 & 0.1 & 75 & 3.4 & 5.8 & 70.7 & $1100$ & $\gamma(\Sigma)$ EOS  \\
		\hline
		EOS$\mathrm{\Sigma}$\_hiMach & 0.09 & 0.004 & 10 & 1.88 & 21.7 & 224 & $6200$ & $\gamma(\Sigma)$ EOS  \\
		\hline
		EOS$\mathrm{\Sigma}$\_ULIRG & 0.42 & 0.0026 & 75 & 0.57 & 14.75 & 2480 & $2.3\times 10^{5}$ & $\gamma(\Sigma)$ EOS  \\
		\hline
		\hline
		EOS$\mathrm{\rho}$\_T10 & 2.3 & 0.1 & 10 & 9.3 & 9.6 & 9.3 & $50$ & $\gamma(\rho)$ EOS  \\
		\hline
		EOS$\mathrm{\rho}$\_T20 & 4.6 & 0.1 & 20 & 6.5 & 8.1 & 18.6 & $150$ & $\gamma(\rho)$ EOS  \\
		\hline
		EOS$\mathrm{\rho}$\_T75 & 17.25 & 0.1 & 75 & 3.4 & 5.8 & 70.7 & $1100$ & $\gamma(\rho)$ EOS  \\
		\hline
		EOS$\mathrm{\rho}$\_hiMach & 0.09 & 0.004 & 10 & 1.88 & 21.7 & 224 & $6200$ &   $\gamma(\rho)$ EOS  \\
		\hline
		EOS$\mathrm{\rho}$\_ULIRG  & 0.42 & 0.0026 & 75 & 0.57 & 14.75 & 2480 & $2.3\times 10^{5}$ & $\gamma(\rho)$ EOS  \\
		\hline
		\hline
		EOSPhys\_T10  & 2.3 & 0.1 & 10 & 9.3 & 9.6 & 9.3 & $50$ & $\gamma_{\rm phys}(\rho)$ EOS  \\
		\hline
		\hline
		Heated\_T10 & 2.3 & 0.1 & 10 & 9.3 & 9.6 & 9.3 & $50$ & Protostellar Heating  \\
		\hline
		Heated\_T20 & 4.6 & 0.1 & 20 & 6.5 & 8.1 & 18.6 & $150$ & Protostellar Heating  \\
		\hline
		Heated\_T75 & 17.25 & 0.1 & 75 & 3.4 & 5.8 & 70.7 & $1100$ & Protostellar Heating  \\
		\hline
		Heated\_hiMach & 0.09 & 0.004 & 10 & 1.88 & 21.7 & 224 & $6200$ &  Protostellar Heating  \\
		\hline
		Heated\_hiDens & 0.46 & 0.01 & 20 & 2.08 & 14.4 & 183 & $5400$ & Protostellar Heating  \\
		\hline
		Heated\_ULIRG  & 0.42 & 0.0026 & 75 & 0.57 & 14.8 & 2480 & $2.3\times 10^{5}$ & Protostellar Heating  \\
		\hline
		\end{tabular}
	\caption{Initial conditions of the different simulation runs presented in this paper. The actual input parameters of the code are the sonic mass $M_{\rm sonic}$ and length $R_{\rm sonic}$, from which more physical parameters like initial temperature ($T_0$), radius ($R$), Mach number ($\mach_0$), surface density ($\Sigma_0$) and number density ($n_{\rm{H}_2,0}$) can be derived. All runs were performed for a large statistical ensemble ($\sim 500$) of $10^4\,\solarmass$ GMCs.}
	\label{tab:simparam}
\end{table*}

\section{Source of Invariant Mass Scale}\label{sec:inv_mass_scale}

One of the key features of the IMF is the turnover mass which appears to be close to universal. In this section we investigate different models of turbulent fragmentation -- starting from the simplest -- to test whether they are capable of producing a nearly-invariant turnover mass, as demanded by the observations.

\subsection{Failure of Isothermal Fragmentation}\label{sec:isothermal}

\begin{figure}
\begin {center}
\includegraphics[width=\linewidth]{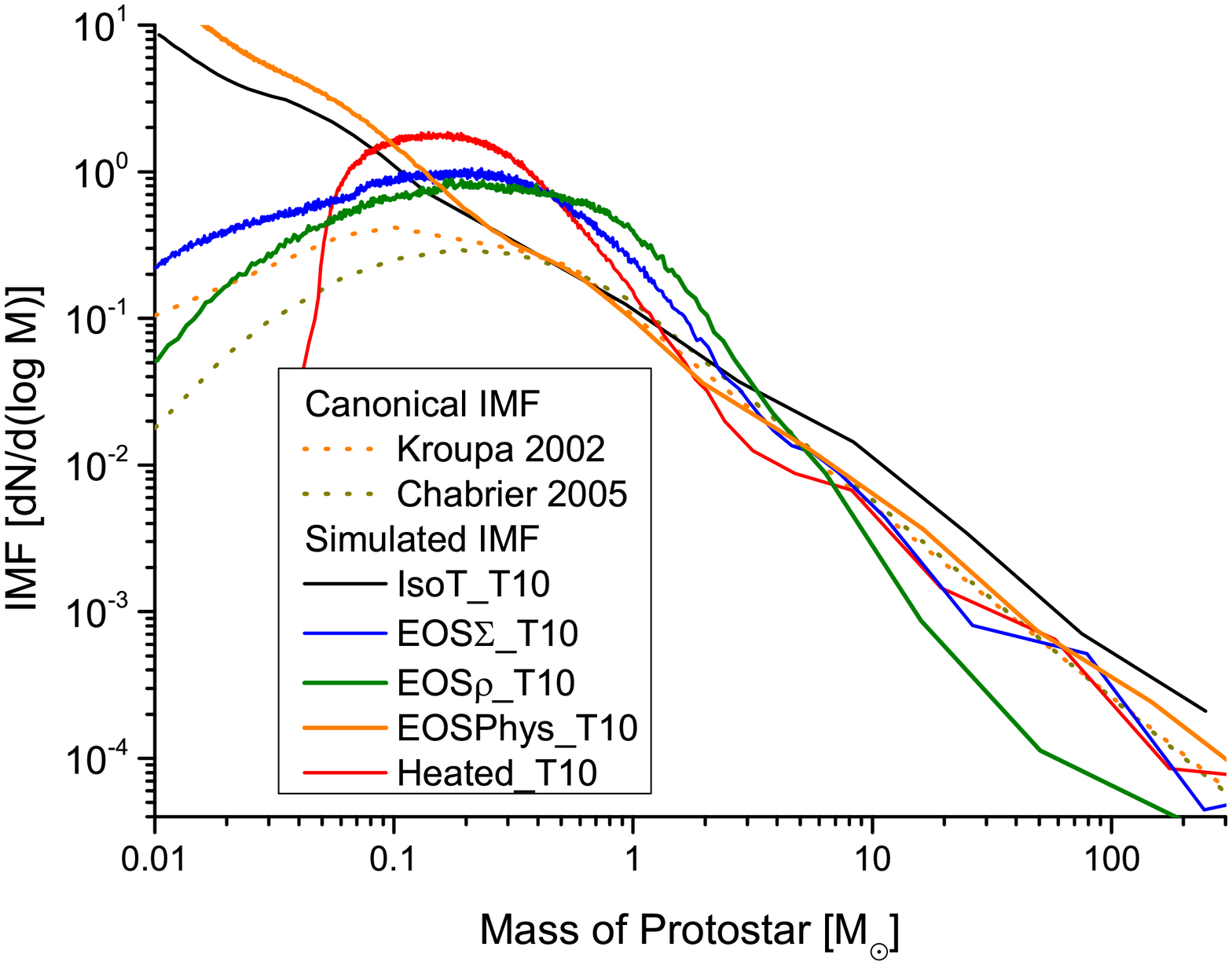}
\caption{The IMF in the case of purely isothermal equation of state (model \textit{IsoTherm\_T10},  solid black), a surface density dependent ``stiff'' EOS (\textit{EOS$\mathrm{\Sigma}$\_T10}, solid blue), a volume density dependent ``stiff'' EOS (\textit{EOS$\mathrm{\rho}$\_T10}, solid green), a physically motivated ``stiff'' EOS (\textit{EOSPhys\_T10}, solid orange) and a protostellar heating (\textit{Heating\_T10}, solid red) model. We compare these to the canonical IMFs of \protect\cite{Kroupa_IMF} and \protect\cite{Chabrier_IMF}. Isothermal collapse leads to a featureless power-law close to $\dderiv N/\dderiv M\propto M^{-2}$ while both protostellar heating and the EOS introduce a turnover at lower masses while having close to canonical behavior at higher masses. Although the physically motivated EOS of Eq. \ref{eq:gamma_phys_rho} does create a turnover, it is at such a low mass that the resulting IMF looks like a power law in the stellar mass range.}
\label{fig:PSMF_canonical_compare}
\end {center}
\end{figure}

We first examine our isothermal case, \textit{IsoTherm\_T10}, the results for which are shown in Fig. \ref{fig:PSMF_canonical_compare}. As the Figure shows, the IMF we obtain in the isothermal case is a pure power-law, with no visible turnover. Although not shown in Figure \ref{fig:PSMF_canonical_compare}, we obtain a similar scale-free result for the IMF produced by purely isothermal fragmentation independent of our choice of initial conditions. It is important to note that, in the isothermal case, the core mass function (CMF) does have a turnover, at the sonic mass $M_{\rm sonic}\sim c_s^2 R_{\rm sonic}/G$, which is set by the initial conditions (see \framework). However, this does not result in an IMF with a turnover.

This result might at first seem surprising, but we can understand it through a simple analytic argument. In a number of analytical studies (e.g. \citealt{core_IMF}) the IMF is inferred from the CMF by shifting the mass scale by a factor of 1/3 (rule of thumb: ``a third of the bound mass ends up in the star''), which is not physically correct, as cores undergo gravitational collapse which takes a finite amount of time, allowing them to further fragment into a spectrum of submasses (\citealt{GuszejnovIMF}). 

This means that a single initial core forms its own sub-cores starting from different initial conditions, so the distribution of subfragments (CMF of \myquote{second generation} fragments) will have its turnover at a different scale than the parent population. The collapse of highly supersonic clouds is self-similar so every factor of 2 contraction takes about a dynamical time (see Sec. 9.2 in \cite{general_turbulent_fragment}). This means that the cloud can fragment at any scale thus there is the same ``amount of fragmentation'' at each scale, producing an infinite fragmentation cascade\footnote{Note that to have finite mass in any mass bins the cascade can not be infinite, it has to be terminated at some finite scale by additional physics. So our isothermal model still has two mass scales: 1) the outer scale set by initial conditions (e.g. GMC mass); 2) the cascade termination scale. If these scales are sufficiently far, a scale-free regime forms between them (similar to the inertial range in turbulence). Assuming that the stellar mass scale is much higher than the termination scale, the distribution in the that range must be close to the self similar.}. This explains why numerical studies have been unable to get converged results, as higher resolution leads to fragmentation on even smaller scales.

We will now attempt to illustrate the qualitative behavior we might expect from a self similar fragmentation cascade, by calculating the IMF in a special case. First, let us assume that a self gravitating cloud has $\lambda$ chance of collapsing without fragmentation and forming a star. Because the process is self-similar, $\lambda$ must be independent of cloud mass. Let us further assume that when a cloud of $M$ mass fragments, the newly formed clouds have an average mass of $\alpha M$. For convenience let us further simplify the model by assuming that a cloud either collapses to a star or breaks up into fragments of $\alpha$ relative mass.


In this simplified model, calculating the mass budget is very easy. The $i$th generation of fragmentation produces clouds of mass $M_i=\alpha^i M_0$, where $M_0$ is the mass of the initial cloud. The total mass of these clouds is $M_0 (1-\lambda)^i$, where the second factor is simply the fraction of the mass not collapsed to stars yet in the previous $i-1$ generations. Since a fraction $\lambda$ of these clouds will collapse to stars without fragmenting further, the total mass of stars of mass $M_i$ is just $f_i = M_0 \lambda (1-\lambda)^i$. As mentioned in Sec.~1, the results from numerical simulations show a large degree of fragmentation, so we expect $\lambda \ll 1$. In this limit, $f_i \approx M_0 \lambda (1 - i \lambda)$, and $f_i$ will therefore be approximately constant for all $i \ll 1/\lambda$. Since $1/\lambda \gg 1$, this means that $f_i$ is nearly constant over a very large number of generations of fragmentation. Further recall that, since the generations of of fragments are separated logarithmically in mass (i.e., $\log (M_i/M_{i+1}) = \log\alpha$ is constant), a constant value of $f_i$ corresponds to constant mass per logarithmic interval in object mass. In terms of number of objects per unit mass (as opposed to per unit log mass), this is $dN/dM \propto M^{-2}$, which is close to what we find. Our actual model is considerably more complex, in that clouds can produce variable numbers of fragments with variable masses, but this simple illustration captures the essence of the isothermal result.

In summary: although isothermal models like \cite{core_IMF} recover the CMF shape, they are unable to explain the shape of the IMF. In the case of isothermal fragmentation, independent of the form of the CMF, the IMF becomes a power-law of $M^{-2}$ as the initial conditions are ``forgotten'' during the fragmentation cascade. This means that to produce an IMF that is not a pure power-law, as observed, an extra physical process is required that would stop the cascade at a mass scale invariant to the initial conditions.

\subsection{Can a Universal Mass Scale Come from the Equation of State?}\label{sec:EOS}

One mechanism to imprint a mass scale onto the process of turbulent fragmentation is to have the equation of state deviate from isothermality, either because the gas becomes optically thick to its own cooling radiation, or due to a change in the cooling process such as the onset of grain-gas coupling. We investigate this approach in our EOS models.

\begin{figure}
\begin {center}
\includegraphics[width=\linewidth]{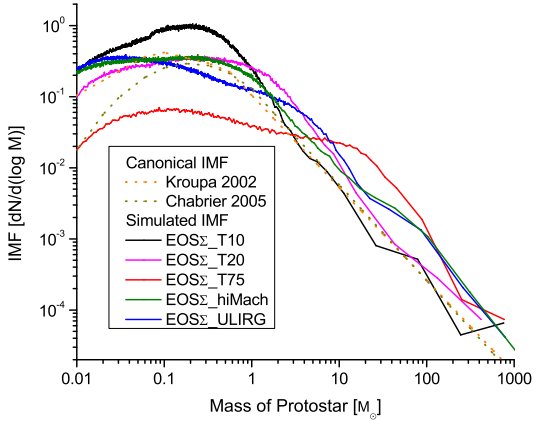}
\caption{The IMF of the surface density dependent EOS model (\textit{EOS$\mathrm{\Sigma}$}) for standard (\textit{EOS$\mathrm{\Sigma}$\_T10}: $T=10\,\kelvin$, $R_{\rm sonic}=0.1\,\pc$), high temperature (\textit{EOS$\mathrm{\Sigma}$\_T20}: $T=20\,\kelvin$),  extreme turbulence (\textit{EOS$\mathrm{\Sigma}$\_hiMach}: $R_{\rm sonic}=0.0026\,\pc$), extreme temperature (\textit{EOS$\mathrm{\Sigma}$\_T75}: $T=75\,\kelvin$) and ULIRG (\textit{EOS$\mathrm{\Sigma}$\_ULIRG}: $T=75\,\kelvin, R_{\rm sonic}=0.0026\,\pc$) initial conditions (see Table \ref{tab:simparam}). There is a clear trend of increasing turnover mass with initial temperature, consistent with our expectation that, for these EOS models, the turnover should scale as $M_{\rm crit}\propto T_0^2$.}
\label{fig:EOS_compare}
\end {center}
\end{figure}

Figure \ref{fig:EOS_compare} shows the results of simulations using our $\gamma(\Sigma)$ (surface density-dependent) EOS (\textit{EOS$\mathrm{\Sigma}$} models), for a variety of initial conditions. We see that, with an appropriate choice of $\Sigma_{\rm crit}$, one can obtain a stellar mass function that agrees reasonably well with the observed IMF. However, one can do so only for a particular choice of initial conditions. As shown in \framework, an EOS with stiffening suppresses fragmentation below mass scale $M_{\rm crit}\sim \frac{c_s^4}{\Sigma_{\rm crit} G^2}\propto T^2/\Sigma_{\rm crit}$ which is clearly shown by the Figure. Also, stronger turbulence leads to more fragmentation and thus more brown dwarfs (see \textit{EOS$\mathrm{\Sigma}$\_T75} and \textit{EOS$\mathrm{\Sigma}$\_ULIRG}) in accordance with predictions (e.g. \protect\citealt{Hopkins_CMF_var}). At first \textit{EOS$\mathrm{\Sigma}$\_hiMach} might seem to contradict that as it has more large protostars than the standard case. This, however, is caused by the interaction of the initial conditions with the adopted EOS. In this scenario the initial surface density $\Sigma_{\rm init}\sim M_{\rm sonic}/\left(R_{\rm sonic}^2 8\pi\right)\propto T/R_{\rm sonic}$ is already above the stiffening transition surface density $\Sigma_{\mathrm{crit}}$. As a result, there is very little fragmentation because the EOS is always ``stiff''. It is also worth noting that the EOS model always has a slow cutoff at low masses despite the fact that protostellar disk fragmentation (a potential source of brown dwarfs) is neglected, so it is likely to overproduce brown dwarfs,  

\begin{figure}
\begin {center}
\includegraphics[width=\linewidth]{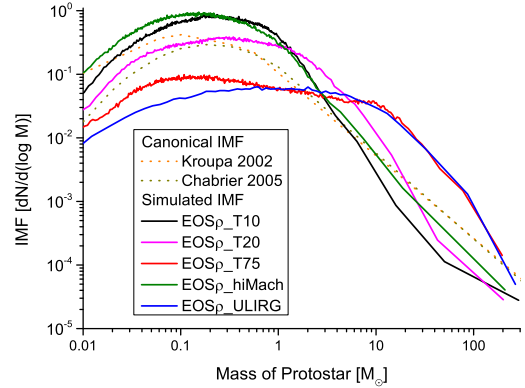}
\caption{The IMF of the volume density dependent EOS model (\textit{EOS$\mathrm{\rho}$})for standard (\textit{EOS$\mathrm{\rho}$\_T10}: $T=10\,\kelvin$, $R_{\rm sonic}=0.1\,\pc$), high temperature (\textit{EOS$\mathrm{\rho}$\_T20}: $T=20\,\kelvin$),  extreme turbulence (\textit{EOS$\mathrm{\rho}$\_hiMach}: $R_{\rm sonic}=0.0026\,\pc$), extreme temperature (\textit{EOS$\mathrm{\rho}$\_T75}: $T=75\,\kelvin$) and ULIRG (\textit{EOS$\mathrm{\rho}$\_ULIRG}: $T=75\,\kelvin, R_{\rm sonic}=0.0026\,\pc$) initial conditions (see Table \ref{tab:simparam}). There is a clear trend of increasing turnover mass with initial temperature, consistent with $M_{\rm crit}\propto T_0^{3/2}$. Despite having stronger turbulence \textit{EOS$\mathrm{\rho}$\_ULIRG} seems to produce more top heavy IMF than \textit{EOS$\mathrm{\rho}$\_T75}. This occurs because in this model the initial density starts out very close to the critical density.}
\label{fig:EOS_rho_compare}
\end {center}
\end{figure}

We have similarly tested an EOS that becomes stiff at a critical volume density $\rho_{\rm crit}$ (see Eq. \ref{eq:gamma_rho}). Fig. \ref{fig:EOS_rho_compare} shows that, as in the case for the $\gamma(\Sigma)$ models, the volume density dependent EOS is also very sensitive to initial conditions. This can be easily understood using a similar arguments as the ones used by \framework in the $\gamma(\Sigma)$ case: using the collapse condition and size-mass relations (see Sec. 2.2 in \framework) one can find the size and mass of a self gravitating fragment whose density is $\rho_{\rm crit}$, which leads to the corresponding turnover mass scale $M_{\rm crit}\approx M_{\rm Jeans}(\rho_{\rm crit})\propto T_0^{3/2} \rho_{\rm crit}^{-1/2}$ (this is also shown by \citealt{Bate_2009_rad_importance}).

We have therefore shown that, while it is possible to choose critical values $\Sigma_{\rm crit}$ or $\rho_{\rm crit}$ such that a stiffened equation of state produces an IMF peak that is qualitatively consistent with observations, such a choice works for only one particular set of initial conditions (see Figures \ref{fig:EOS_compare}-\ref{fig:EOS_rho_compare}). Substantially different initial temperatures necessitate different choices to keep the IMF peak fixed, and there is no obvious physical reason why the critical parameters should vary in such a manner. Indeed, we remind readers that even the values we have used for the standard MW case ($T_0 = 10\,\rm{K}$) have been optimized to fit the observations, and are not motivated by any plausible physical model. Choosing the values of $\Sigma_{\rm crit}$ or $\rho_{\rm crit}$ that one would naturally predict based on considerations of gas thermodynamics would make the agreement with observations very poor even in the Milky Way-like case (see \textit{EOSPhys\_T10} in Fig. \ref{fig:PSMF_canonical_compare}).

\subsection{Effects of Protostellar Heating}\label{sec:heating}

Another proposed origin of a universal mass scale is stellar feedback, including protostellar heating, outflows, accretion, photo-ionization heating and supernovae, none of which are scale-free processes. Thus they all have the capability to imprint a mass scale. In this paper we only concentrate on protostellar heating as it is the earliest and strongest feedback mechanism during the evolution of protostellar cores. Most of the other mechanisms act \emph{after} the stars form, which can therefore only alter the IMF of ``second generation'' stars.

\begin{figure}
\begin {center}
\includegraphics[width=\linewidth]{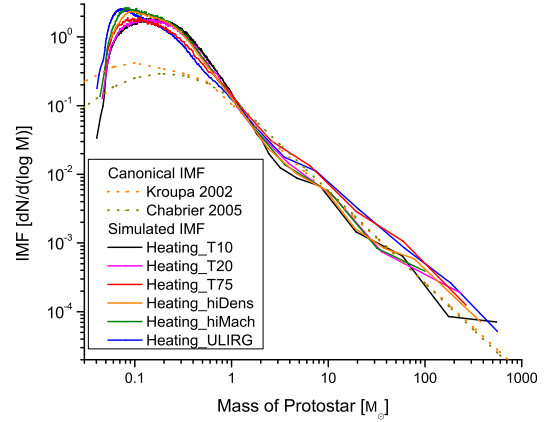}
\caption{The IMF of the protostellar heating model with standard (\textit{Heating\_T10}: $T=10\,\kelvin$, $R_{\rm sonic}=0.1\,\pc$), high temperature (\textit{Heating\_T20}: $T=20\,\kelvin$), high density and temperature (\textit{Heating\_hiDens}: $T=20\,\kelvin$, $n=5000\,\mathrm{cm}^{-3}$),  extreme turbulence (\textit{Heating\_hiMach}: $R_{\rm sonic}=0.0026\,\pc$), extreme temperature (\textit{Heating\_T75}: $T=75\,\kelvin$) and ULIRG (\textit{Heating\_ULIRG}: $T=75\,\kelvin, R_{\rm sonic}=0.0026\,\pc$) initial conditions (see Table \ref{tab:simparam}). The predicted IMF is remarkably invariant to initial conditions. The turnover point does shift slightly to lower masses for both very strong turbulence and high temperature (stronger turbulence makes fragmentation easier and a higher initial temperature means that protostellar heating becomes dominant at a smaller size scale).}
\label{fig:PSMF_initial_conditions}
\end {center}
\end{figure}

Figure \ref{fig:PSMF_initial_conditions} shows the results of our calculation including protostellar heating. Similar to the EOS models, at very high masses the additional physics (protostellar heating) has no significant effect, and thus the IMF looks similar to the isothermal result of $M^{-2}$. The isothermal fragmentation cascade is terminated around the characteristic mass of the model, creating a ``pile up''. Note that the current model underproduces brown dwarfs as it neglects disk fragmentation, and more generally any fragmentation process that depends on angular momentum. As the figure shows, inclusion of heating produces a peak that is consistent with the observed peak of the IMF, and that is remarkably insensitive to changes in the star-forming environment. The only changes in the position of the peak visible in Figure \ref{fig:PSMF_initial_conditions} are in the \textit{ULIRG} and \textit{hiMach} runs, where the peak is shifted to lower masses by a factor of $\sim 2$. The \textit{hiDens} run, which is set up to emulate a dense star forming region in the Milky Way, is intermediate between these two cases and the normal Milky Way case, with a peak that is shifted by a tens of percent slightly relative to \textit{T10}. We emphasize that, unlike the $\gamma(\Sigma)$ and $\gamma(\rho)$ cases where we explicitly tuned model parameters to produce the correct peak mass, the protostellar heating model is not tuned, and has no free parameters. Its only parameter is the value of $\Psi$, which is determined entirely by the physics of stellar structure and deuterium burning. Thus both the location and the invariance of the IMF peak in this model are independent predictions.

It is worth noting that this model does seem to produce too few brown dwarfs and an excess of M dwarf stars. However, it also neglects protostellar disk fragmentation and other ``sources'' of brown dwarfs, which would reduce the excess between $0.1-1\,\solarmass$ and enhance the number of objects at lower masses. Whether including these processes leads to the correct proportion of brown dwarfs remains an open question, though the radiation-hydrodynamic simulations of \cite{Bate09a, Bate2014} and \cite{Krumholz12a} suggest this is in fact the case.

It is also instructive to compare the results of the protostellar heating models to the EOS models, in order to understand why the results are so different. We use a simple model that assumes the cloud behaves ``isothermally'' except for a global heating term. This means that $T_{\rm EOS}=T_0$ (from Eq. \ref{eq:T_summing}), which is the initial temperature of the cloud (set by external heating like cosmic rays). At first glance the protostellar heating model proposed above seems very much like an opacity limit EOS model, as $T_{\rm heat}\propto M^{3/8}R^{-7/8}\approx\Sigma^{3/8}$ so the collapse of the cloud is isothermal until a characteristic $\Sigma_{\rm heat}$ is reached where $T_{\rm heat}=T_0$. From that point on $T\approx T_{\rm heat}$ which means that the temperature increases as if we had a polytropic index of $\gamma=31/24$ (see Eq. \ref{eq:gamma}). Similar to the EOS models we can find the characteristic fragment mass $M_{\rm crit}$ where this transition happens. Using the above relations, the collapse threshold $\frac{M}{R}\frac{1}{T}\frac{1}{1+\mach^2}=\rm{const.}$ and assuming a subsonic fragment ($\mach\ll 1$) we get $M_{\rm crit} T_0^{1/4}=\rm{const.}$ which means that there is remarkably weak sensitivity to the initial temperature (\Markfeedback includes a more rigorous derivation which yields $M_{\rm crit}\propto T_0^{-1/18}$). Comparing Fig. \ref{fig:PSMF_initial_conditions} with Fig. \ref{fig:EOS_compare} makes the difference this produces in the resulting IMF abundantly clear.

\section{Conclusions}\label{sec:conclusions}

The aim of this paper is to investigate what physical processes can explain the origin of the IMF, and in particular the fact that the IMF is not a powerlaw, and that its characteristic mass scale is remarkably insensitive to variations in the star-forming environment. To this end, we have considered three classes of models for gas thermodynamics: purely isothermal models, models with an equation of state that stiffens at a characteristic volume or surface density, and models containing a simple analytic estimate for the effects of protostellar heating. 

We find that purely isothermal models categorically fail to reproduce the IMF. Although the initial conditions do imprint a mass scale (the sonic mass) which is apparent in the distribution of bound structures (i.e., the CMF), due to the lack of mass scale in the equations of motion this scale is ``forgotten'' during the fragmentation cascade, leading to an $M^{-2}$ power-law solution for the IMF (consistent with the lack of convergence reported thus far in numerical studies). This means that isothermal gravito-turbulent fragmentation cannot explain the existence or universality of the turnover scale in the IMF. Some other physics is needed for that.

An often invoked expansion of the fragmentation model is to have the clouds transition from an isothermal to a ``stiff'' equation of state when they reach a critical surface or volume density and become thick to their own cooling radiation. This does provide a mass scale for the system, and by tuning the parameters of the model appropriately one can reproduce the observed IMF turnover. However, we find that this approach results in a mass scale that is extremely sensitive to initial conditions ($M_{\rm crit}[\gamma\left(\Sigma\right)]\propto T^2$ and $M_{\rm crit}[\gamma\left(\rho\right)]\propto T^{3/2}$), rendering these models unable to provide a universal mass scale as is observed. Moreover, producing agreement with the observed mass scale even for initial conditions similar to those found in Solar neighborhood star-forming regions requires parameter choices that are very far from what one would have estimated based on any first-principles physical argument.

We argue instead that feedback physics can provide a mass scale that is both in good agreement with observations and insensitive to the conditions in the star-forming region. As an example, based on \cite{Krumholz_stellar_mass_origin}, we have formulated a simple prescription for protostellar heating. This alone of all the analytical models we consider is able to provide a \emph{universal} IMF turnover, despite large variations in initial gas temperature, densities, Mach number and masses of star forming clouds.

\subsection{Caveats and Future Work}

We close with a discussion of the limitations of our model, and how we plan to improve it in future work. We utilize the semi-analytical framework of \cite{TurbFramework} which makes strong approximations. Motivated by \cite{Brant_turb_pumping} we assume collapse at constant virial parameter as turbulence is pumped by gravity. While this assumption has empirical support, it has not been rigorously demonstrated (although simulations so far seem to confirm this, see \citealt{Murray_2015_turb_sim}). Furthermore, the simulation only follows the evolution of self gravitating structures until they reach the size scale where angular momentum becomes important (which is not treated in the current models), and, thus processes that act on the scales of disks or smaller (e.g. disk fragmentation) are neglected. This could have a significant effect on the low mass end of the resulting IMF. Also, fragments are assumed to evolve independently, so mergers and other interactions are neglected\footnote{This is actually a fairly good assumption. The timescale for two clouds of $R$ radius to merge in this framework is $t_{\rm merger}\sim d/v$, where $d$ is the separation between clouds and $v$ is their relative velocity towards each other. It is easy to show that $t_{\rm merger}/t_{\rm freefall}\sim \sqrt{d/R\left(1+R_{\rm sonic}/R\right)}>\sqrt{d/R}$. This means that the timescale for merging is only comparable to the freefall time if the clouds initially form right next to each other ($d\sim 2 R$).}. Finally, the protostellar heating model assumes instantenous, isotropic, steady state heating and neglects other forms of feedback (e.g. outflows).  

Some of these limitations will be easier to remove than others. The assumption that collapse occurs at constant virial parameter can be investigated by simulations, as can be the fragmentation of disks, and in principle results from these calculations could be incorporated into our model. Similarly, a number of authors have proposed more complex models for the protostellar heating, including the effects of fluctuations in time (e.g., \citealt{Lomax_heating_fluct_sim}) that was found to have significant effect on the statistics of star formation (\citealt{Stamatellos_radfeedback,Lomax_multiplicity}), and these could be included as well. Furthermore, it is possible to include angular momentum (like in \citealt{core_IMF}) and interaction between fragments with significant extension of the model. The entire framework can also be checked against radiation-hydrodynamic simulations such as those of \cite{Krumholz12a} or \cite{Myers2014_sim}. 

In addition to these improvements in the model itself, an obvious next step is to identify predictions of the model that can be compared with real data. We mention here two obvious, first order predictions that we plan to investigate in future work. First, using the output of cosmological simulations or semi-analytic models, we can investigate the extent to which the small amount of variation we do find in the protostellar heating model produces significant variations in the IMFs of elliptical galaxies over cosmological times. These predictions can then be compared to observations (e.g., \citealt{Conroy_vanDokkum_ellipticals, Cappellari_IMF_var_2012}). Second, because our model retains spatial information, it makes predictions for the clustering of stars as well as for their mass distribution. This too can be checked against the spatial distribution of stars in nearby star-forming regions, a test that has been performed before using both analytic (\citealt{Hopkins_clustering}) and numerical (\citealt{Hansen_lowmass_SF_feedback, Myers2014_sim}) models. It should be noted however, that without accounting for protostellar disk fragmentation most results (e.g. correlation function, binarity) will only be valid on scales larger than the typical protostellar disk size.

\acknowledgments
Support for PFH and DG was provided by an Alfred P. Sloan Research Fellowship, NASA ATP Grant NNX14AH35G, and NSF Collaborative Research Grant \#1411920 and CAREER grant \#1455342. MRK is supported by NSF grants AST-0955300, AST-1405962 and NASA ATP grant NNX15AT06G. Numerical calculations were run on the Caltech computer cluster ``Zwicky'' (NSF MRI award \#PHY-0960291) and allocation TG-AST130039 granted by the Extreme Science and Engineering Discovery Environment (XSEDE) supported by the NSF. 

\bibliographystyle{mnras}
\bibliography{bibliography}

\begin{thebibliography}{}
\makeatletter
\relax
\def\mn@urlcharsother{\let\do\@makeother \do\$\do\&\do\#\do\^\do\_\do\%\do\~}
\def\mn@doi{\begingroup\mn@urlcharsother \@ifnextchar [ {\mn@doi@}
  {\mn@doi@[]}}
\def\mn@doi@[#1]#2{\def\@tempa{#1}\ifx\@tempa\@empty \href
  {http://dx.doi.org/#2} {doi:#2}\else \href {http://dx.doi.org/#2} {#1}\fi
  \endgroup}
\def\mn@eprint#1#2{\mn@eprint@#1:#2::\@nil}
\def\mn@eprint@arXiv#1{\href {http://arxiv.org/abs/#1} {{\tt arXiv:#1}}}
\def\mn@eprint@dblp#1{\href {http://dblp.uni-trier.de/rec/bibtex/#1.xml}
  {dblp:#1}}
\def\mn@eprint@#1:#2:#3:#4\@nil{\def\@tempa {#1}\def\@tempb {#2}\def\@tempc
  {#3}\ifx \@tempc \@empty \let \@tempc \@tempb \let \@tempb \@tempa \fi \ifx
  \@tempb \@empty \def\@tempb {arXiv}\fi \@ifundefined
  {mn@eprint@\@tempb}{\@tempb:\@tempc}{\expandafter \expandafter \csname
  mn@eprint@\@tempb\endcsname \expandafter{\@tempc}}}

\bibitem[\protect\citeauthoryear{{Alves}, {Lombardi}  \& {Lada}}{{Alves}
  et~al.}{2007}]{Alves_CMF_IMF_obs}
{Alves} J.,  {Lombardi} M.,   {Lada} C.~J.,  2007, \mn@doi [\aap]
  {10.1051/0004-6361:20066389}, \href
  {http://adsabs.harvard.edu/abs/2007A%26A...462L..17A} {462, L17}

\bibitem[\protect\citeauthoryear{{Bastian}, {Covey}  \& {Meyer}}{{Bastian}
  et~al.}{2010}]{IMF_review}
{Bastian} N.,  {Covey} K.~R.,   {Meyer} M.~R.,  2010, \mn@doi [\araa]
  {10.1146/annurev-astro-082708-101642}, \href
  {http://adsabs.harvard.edu/abs/2010ARA%26A..48..339B} {48, 339}

\bibitem[\protect\citeauthoryear{{Bate}}{{Bate}}{2009a}]{Bate09a}
{Bate} M.~R.,  2009a, \mnras, 392, 590

\bibitem[\protect\citeauthoryear{{Bate}}{{Bate}}{2009b}]{Bate_2009_rad_importance}
{Bate} M.~R.,  2009b, \mn@doi [\mnras] {10.1111/j.1365-2966.2008.14165.x},
  \href {http://adsabs.harvard.edu/abs/2009MNRAS.392.1363B} {392, 1363}

\bibitem[\protect\citeauthoryear{{Bate}}{{Bate}}{2012}]{Bate12a}
{Bate} M.~R.,  2012, \mnras, 419, 3115

\bibitem[\protect\citeauthoryear{{Bate}}{{Bate}}{2014}]{Bate2014}
{Bate} M.~R.,  2014, \mnras, 442, 285

\bibitem[\protect\citeauthoryear{{Bolatto}, {Leroy}, {Rosolowsky}, {Walter}  \&
  {Blitz}}{{Bolatto} et~al.}{2008}]{Bolatto_2008}
{Bolatto} A.~D.,  {Leroy} A.~K.,  {Rosolowsky} E.,  {Walter} F.,   {Blitz} L.,
  2008, \mn@doi [\apj] {10.1086/591513}, \href
  {http://adsabs.harvard.edu/abs/2008ApJ...686..948B} {686, 948}

\bibitem[\protect\citeauthoryear{{Bonnell}, {Clarke}  \& {Bate}}{{Bonnell}
  et~al.}{2006}]{Bonnell06a}
{Bonnell} I.~A.,  {Clarke} C.~J.,   {Bate} M.~R.,  2006, \mnras, 368, 1296

\bibitem[\protect\citeauthoryear{{Cappellari} et~al.,}{{Cappellari}
  et~al.}{2012}]{Cappellari_IMF_var_2012}
{Cappellari} M.,  et~al., 2012, \mn@doi [\nat] {10.1038/nature10972}, \href
  {http://adsabs.harvard.edu/abs/2012Natur.484..485C} {484, 485}

\bibitem[\protect\citeauthoryear{{Chabrier}}{{Chabrier}}{2005}]{Chabrier_IMF}
{Chabrier} G.,  2005, in {Corbelli} E.,  {Palla} F.,   {Zinnecker} H.,  eds,
  Astrophysics and Space Science Library Vol. 327, The Initial Mass Function 50
  Years Later. p.~41

\bibitem[\protect\citeauthoryear{{Dobbs}, {Bonnell}  \& {Pringle}}{{Dobbs}
  et~al.}{2006}]{Dobbs06a}
{Dobbs} C.~L.,  {Bonnell} I.~A.,   {Pringle} J.~E.,  2006, \mnras, 371, 1663

\bibitem[\protect\citeauthoryear{{Elmegreen}, {Klessen}  \&
  {Wilson}}{{Elmegreen} et~al.}{2008}]{Elmegreen08a}
{Elmegreen} B.~G.,  {Klessen} R.~S.,   {Wilson} C.~D.,  2008, \apj, 681, 365

\bibitem[\protect\citeauthoryear{{Geha} et~al.,}{{Geha}
  et~al.}{2013}]{Geha_turnover_univ}
{Geha} M.,  et~al., 2013, \mn@doi [\apj] {10.1088/0004-637X/771/1/29}, \href
  {http://adsabs.harvard.edu/abs/2013ApJ...771...29G} {771, 29}

\bibitem[\protect\citeauthoryear{{Glover} \& {Mac Low}}{{Glover} \& {Mac
  Low}}{2007}]{Glover_EQS_lowgamma_ref}
{Glover} S.~C.~O.,  {Mac Low} M.-M.,  2007, \mn@doi [\apjs] {10.1086/512238},
  \href {http://adsabs.harvard.edu/abs/2007ApJS..169..239G} {169, 239}

\bibitem[\protect\citeauthoryear{{Goodwin}, {Whitworth}  \&
  {Ward-Thompson}}{{Goodwin} et~al.}{2004}]{Goodwin04a}
{Goodwin} S.~P.,  {Whitworth} A.~P.,   {Ward-Thompson} D.,  2004, \aap, 414,
  633

\bibitem[\protect\citeauthoryear{{Guszejnov} \& {Hopkins}}{{Guszejnov} \&
  {Hopkins}}{2015a}]{TurbFramework}
{Guszejnov} D.,  {Hopkins} P.~F.,  2015a, preprint, \href
  {http://adsabs.harvard.edu/abs/2015arXiv150706678G} {} (\mn@eprint {arXiv}
  {1507.06678})

\bibitem[\protect\citeauthoryear{Guszejnov \& Hopkins}{Guszejnov \&
  Hopkins}{2015b}]{GuszejnovIMF}
Guszejnov D.,  Hopkins P.~F.,  2015b, \mn@doi [\mnras] {10.1093/mnras/stv872},
  450, 4137

\bibitem[\protect\citeauthoryear{{Hansen}, {Klein}, {McKee}  \&
  {Fisher}}{{Hansen} et~al.}{2012}]{Hansen_lowmass_SF_feedback}
{Hansen} C.~E.,  {Klein} R.~I.,  {McKee} C.~F.,   {Fisher} R.~T.,  2012,
  \mn@doi [\apj] {10.1088/0004-637X/747/1/22}, \href
  {http://adsabs.harvard.edu/abs/2012ApJ...747...22H} {747, 22}

\bibitem[\protect\citeauthoryear{{Hennebelle} \& {Chabrier}}{{Hennebelle} \&
  {Chabrier}}{2008}]{HC08}
{Hennebelle} P.,  {Chabrier} G.,  2008, \mn@doi [\apj] {10.1086/589916}, \href
  {http://adsabs.harvard.edu/abs/2008ApJ...684..395H} {684, 395}

\bibitem[\protect\citeauthoryear{{Hennebelle} \& {Chabrier}}{{Hennebelle} \&
  {Chabrier}}{2009}]{HC2009}
{Hennebelle} P.,  {Chabrier} G.,  2009, \mn@doi [\apj]
  {10.1088/0004-637X/702/2/1428}, \href
  {http://adsabs.harvard.edu/abs/2009ApJ...702.1428H} {702, 1428}

\bibitem[\protect\citeauthoryear{{Hennebelle} \& {Chabrier}}{{Hennebelle} \&
  {Chabrier}}{2013}]{HC_2013}
{Hennebelle} P.,  {Chabrier} G.,  2013, \mn@doi [\apj]
  {10.1088/0004-637X/770/2/150}, \href
  {http://adsabs.harvard.edu/abs/2013ApJ...770..150H} {770, 150}

\bibitem[\protect\citeauthoryear{{Hopkins}}{{Hopkins}}{2012a}]{excursion_set_ism}
{Hopkins} P.~F.,  2012a, \mn@doi [\mnras] {10.1111/j.1365-2966.2012.20730.x},
  \href {http://adsabs.harvard.edu/abs/2012MNRAS.423.2016H} {423, 2016}

\bibitem[\protect\citeauthoryear{{Hopkins}}{{Hopkins}}{2012b}]{core_IMF}
{Hopkins} P.~F.,  2012b, \mn@doi [\mnras] {10.1111/j.1365-2966.2012.20731.x},
  \href {http://adsabs.harvard.edu/abs/2012MNRAS.423.2037H} {423, 2037}

\bibitem[\protect\citeauthoryear{{Hopkins}}{{Hopkins}}{2013a}]{Hopkins_clustering}
{Hopkins} P.~F.,  2013a, \mn@doi [\mnras] {10.1093/mnras/sts147}, \href
  {http://adsabs.harvard.edu/abs/2013MNRAS.428.1950H} {428, 1950}

\bibitem[\protect\citeauthoryear{{Hopkins}}{{Hopkins}}{2013b}]{general_turbulent_fragment}
{Hopkins} P.~F.,  2013b, \mn@doi [\mnras] {10.1093/mnras/sts704}, \href
  {http://adsabs.harvard.edu/abs/2013MNRAS.430.1653H} {430, 1653}

\bibitem[\protect\citeauthoryear{{Hopkins}}{{Hopkins}}{2013c}]{Hopkins_CMF_var}
{Hopkins} P.~F.,  2013c, \mn@doi [\mnras] {10.1093/mnras/stt713}, \href
  {http://adsabs.harvard.edu/abs/2013MNRAS.433..170H} {433, 170}

\bibitem[\protect\citeauthoryear{{Kratter}, {Matzner}, {Krumholz}  \&
  {Klein}}{{Kratter} et~al.}{2010}]{Kratter10a}
{Kratter} K.~M.,  {Matzner} C.~D.,  {Krumholz} M.~R.,   {Klein} R.~I.,  2010,
  \apj, 708, 1585

\bibitem[\protect\citeauthoryear{{Kritsuk}, {Lee}  \& {Norman}}{{Kritsuk}
  et~al.}{2013}]{Kritsuk_larson_supersonic_origin}
{Kritsuk} A.~G.,  {Lee} C.~T.,   {Norman} M.~L.,  2013, \mn@doi [\mnras]
  {10.1093/mnras/stt1805}, \href
  {http://adsabs.harvard.edu/abs/2013MNRAS.436.3247K} {436, 3247}

\bibitem[\protect\citeauthoryear{{Kroupa}}{{Kroupa}}{2002}]{Kroupa_IMF}
{Kroupa} P.,  2002, \mn@doi [Science] {10.1126/science.1067524}, \href
  {http://adsabs.harvard.edu/abs/2002Sci...295...82K} {295, 82}

\bibitem[\protect\citeauthoryear{{Krumholz}}{{Krumholz}}{2006}]{Krumholz06a}
{Krumholz} M.~R.,  2006, \apjl, 641, L45

\bibitem[\protect\citeauthoryear{{Krumholz}}{{Krumholz}}{2011}]{Krumholz_stellar_mass_origin}
{Krumholz} M.~R.,  2011, \mn@doi [\apj] {10.1088/0004-637X/743/2/110}, \href
  {http://adsabs.harvard.edu/abs/2011ApJ...743..110K} {743, 110}

\bibitem[\protect\citeauthoryear{{Krumholz}}{{Krumholz}}{2014}]{SF_big_problems}
{Krumholz} M.~R.,  2014, \mn@doi [\physrep] {10.1016/j.physrep.2014.02.001},
  \href {http://adsabs.harvard.edu/abs/2014PhR...539...49K} {539, 49}

\bibitem[\protect\citeauthoryear{{Krumholz} \& {McKee}}{{Krumholz} \&
  {McKee}}{2008}]{Krumholz08a}
{Krumholz} M.~R.,  {McKee} C.~F.,  2008, \nat, 451, 1082

\bibitem[\protect\citeauthoryear{{Krumholz}, {McKee}  \& {Klein}}{{Krumholz}
  et~al.}{2005}]{krumholz05a}
{Krumholz} M.~R.,  {McKee} C.~F.,   {Klein} R.~I.,  2005, \mn@doi [\nat]
  {10.1038/nature04280}, 438, 332

\bibitem[\protect\citeauthoryear{{Krumholz}, {Klein}  \& {McKee}}{{Krumholz}
  et~al.}{2007}]{Krumholz07a}
{Krumholz} M.~R.,  {Klein} R.~I.,   {McKee} C.~F.,  2007, \apj, 656, 959

\bibitem[\protect\citeauthoryear{{Krumholz}, {Klein}  \& {McKee}}{{Krumholz}
  et~al.}{2011}]{Krumholz11a}
{Krumholz} M.~R.,  {Klein} R.~I.,   {McKee} C.~F.,  2011, \apj, 740, 74

\bibitem[\protect\citeauthoryear{{Krumholz}, {Klein}  \& {McKee}}{{Krumholz}
  et~al.}{2012}]{Krumholz12a}
{Krumholz} M.~R.,  {Klein} R.~I.,   {McKee} C.~F.,  2012, \apj, 754, 71

\bibitem[\protect\citeauthoryear{{Larson}}{{Larson}}{1981}]{Larson_law}
{Larson} R.~B.,  1981, \mnras, \href
  {http://adsabs.harvard.edu/abs/1981MNRAS.194..809L} {194, 809}

\bibitem[\protect\citeauthoryear{{Larson}}{{Larson}}{2005}]{Larson2005}
{Larson} R.~B.,  2005, \mn@doi [\mnras] {10.1111/j.1365-2966.2005.08881.x},
  \href {http://adsabs.harvard.edu/abs/2005MNRAS.359..211L} {359, 211}

\bibitem[\protect\citeauthoryear{{Lomax}, {Whitworth}, {Hubber}, {Stamatellos}
  \& {Walch}}{{Lomax} et~al.}{2014}]{Lomax_heating_fluct_sim}
{Lomax} O.,  {Whitworth} A.~P.,  {Hubber} D.~A.,  {Stamatellos} D.,   {Walch}
  S.,  2014, \mn@doi [\mnras] {10.1093/mnras/stu177}, \href
  {http://adsabs.harvard.edu/abs/2014MNRAS.439.3039L} {439, 3039}

\bibitem[\protect\citeauthoryear{{Lomax}, {Whitworth}, {Hubber}, {Stamatellos}
  \& {Walch}}{{Lomax} et~al.}{2015}]{Lomax_multiplicity}
{Lomax} O.,  {Whitworth} A.~P.,  {Hubber} D.~A.,  {Stamatellos} D.,   {Walch}
  S.,  2015, \mn@doi [\mnras] {10.1093/mnras/stu2530}, \href
  {http://adsabs.harvard.edu/abs/2015MNRAS.447.1550L} {447, 1550}

\bibitem[\protect\citeauthoryear{{Luhman}, {Mamajek}, {Allen}  \&
  {Cruz}}{{Luhman} et~al.}{2009}]{Luhman_Taurus}
{Luhman} K.~L.,  {Mamajek} E.~E.,  {Allen} P.~R.,   {Cruz} K.~L.,  2009,
  \mn@doi [\apj] {10.1088/0004-637X/703/1/399}, \href
  {http://adsabs.harvard.edu/abs/2009ApJ...703..399L} {703, 399}

\bibitem[\protect\citeauthoryear{{Martel}, {Evans}  \& {Shapiro}}{{Martel}
  et~al.}{2006}]{Martel_numerical_sim_convergence}
{Martel} H.,  {Evans} II N.~J.,   {Shapiro} P.~R.,  2006, \mn@doi [\apjs]
  {10.1086/500090}, \href {http://adsabs.harvard.edu/abs/2006ApJS..163..122M}
  {163, 122}

\bibitem[\protect\citeauthoryear{{Masunaga} \& {Inutsuka}}{{Masunaga} \&
  {Inutsuka}}{2000}]{Masunaga_EQS_highgamma_ref}
{Masunaga} H.,  {Inutsuka} S.-i.,  2000, \mn@doi [\apj] {10.1086/308439}, \href
  {http://adsabs.harvard.edu/abs/2000ApJ...531..350M} {531, 350}

\bibitem[\protect\citeauthoryear{{Masunaga}, {Miyama}  \&
  {Inutsuka}}{{Masunaga} et~al.}{1998}]{Masunaga1998}
{Masunaga} H.,  {Miyama} S.~M.,   {Inutsuka} S.-i.,  1998, \mn@doi [\apj]
  {10.1086/305281}, \href {http://adsabs.harvard.edu/abs/1998ApJ...495..346M}
  {495, 346}

\bibitem[\protect\citeauthoryear{{McKee}, {Li}  \& {Klein}}{{McKee}
  et~al.}{2010}]{McKee_ambipolar_scaling}
{McKee} C.~F.,  {Li} P.~S.,   {Klein} R.~I.,  2010, \mn@doi [\apj]
  {10.1088/0004-637X/720/2/1612}, \href
  {http://adsabs.harvard.edu/abs/2010ApJ...720.1612M} {720, 1612}

\bibitem[\protect\citeauthoryear{{Murray}, {Chang}, {Murray}  \&
  {Pittman}}{{Murray} et~al.}{2015}]{Murray_2015_turb_sim}
{Murray} D.~W.,  {Chang} P.,  {Murray} N.~W.,   {Pittman} J.,  2015, preprint,
  \href {http://adsabs.harvard.edu/abs/2015arXiv150905910M} {} (\mn@eprint
  {arXiv} {1509.05910})

\bibitem[\protect\citeauthoryear{{Myers}, {Krumholz}, {Klein}  \&
  {McKee}}{{Myers} et~al.}{2011}]{Myers11a}
{Myers} A.~T.,  {Krumholz} M.~R.,  {Klein} R.~I.,   {McKee} C.~F.,  2011, \apj,
  735, 49

\bibitem[\protect\citeauthoryear{{Myers}, {Klein}, {Krumholz}  \&
  {McKee}}{{Myers} et~al.}{2014}]{Myers2014_sim}
{Myers} A.~T.,  {Klein} R.~I.,  {Krumholz} M.~R.,   {McKee} C.~F.,  2014,
  \mn@doi [\mnras] {10.1093/mnras/stu190}, \href
  {http://adsabs.harvard.edu/abs/2014MNRAS.439.3420M} {439, 3420}

\bibitem[\protect\citeauthoryear{{Offner}, {Klein}, {McKee}  \&
  {Krumholz}}{{Offner} et~al.}{2009}]{Offner09a}
{Offner} S.~S.~R.,  {Klein} R.~I.,  {McKee} C.~F.,   {Krumholz} M.~R.,  2009,
  \apj, 703, 131

\bibitem[\protect\citeauthoryear{{Offner}, {Clark}, {Hennebelle}, {Bastian},
  {Bate}, {Hopkins}, {Moraux}  \& {Whitworth}}{{Offner}
  et~al.}{2014}]{IMF_universality}
{Offner} S.~S.~R.,  {Clark} P.~C.,  {Hennebelle} P.,  {Bastian} N.,  {Bate}
  M.~R.,  {Hopkins} P.~F.,  {Moraux} E.,   {Whitworth} A.~P.,  2014, \mn@doi
  [Protostars and Planets VI] {10.2458/azu_uapress_9780816531240-ch003}, \href
  {http://adsabs.harvard.edu/abs/2014prpl.conf...53O} {pp 53--75}

\bibitem[\protect\citeauthoryear{{Padoan} \& {Nordlund}}{{Padoan} \&
  {Nordlund}}{2002}]{Padoan_Nordlund_2002_IMF}
{Padoan} P.,  {Nordlund} {\AA}.,  2002, \mn@doi [\apj] {10.1086/341790}, \href
  {http://adsabs.harvard.edu/abs/2002ApJ...576..870P} {576, 870}

\bibitem[\protect\citeauthoryear{{Padoan}, {Nordlund}  \& {Jones}}{{Padoan}
  et~al.}{1997}]{Padoan_theory}
{Padoan} P.,  {Nordlund} A.,   {Jones} B.~J.~T.,  1997, \mnras, \href
  {http://adsabs.harvard.edu/abs/1997MNRAS.288..145P} {288, 145}

\bibitem[\protect\citeauthoryear{{Press} \& {Schechter}}{{Press} \&
  {Schechter}}{1974}]{PressSchechter}
{Press} W.~H.,  {Schechter} P.,  1974, \mn@doi [\apj] {10.1086/152650}, \href
  {http://adsabs.harvard.edu/abs/1974ApJ...187..425P} {187, 425}

\bibitem[\protect\citeauthoryear{{Rathborne}, {Lada}, {Muench}, {Alves},
  {Kainulainen}  \& {Lombardi}}{{Rathborne} et~al.}{2009}]{Rathborne_CMF}
{Rathborne} J.~M.,  {Lada} C.~J.,  {Muench} A.~A.,  {Alves} J.~F.,
  {Kainulainen} J.,   {Lombardi} M.,  2009, \mn@doi [\apj]
  {10.1088/0004-637X/699/1/742}, \href
  {http://adsabs.harvard.edu/abs/2009ApJ...699..742R} {699, 742}

\bibitem[\protect\citeauthoryear{{Rees}}{{Rees}}{1976}]{Rees1976}
{Rees} M.~J.,  1976, \mnras, \href
  {http://adsabs.harvard.edu/abs/1976MNRAS.176..483R} {176, 483}

\bibitem[\protect\citeauthoryear{{Robertson} \& {Goldreich}}{{Robertson} \&
  {Goldreich}}{2012}]{Brant_turb_pumping}
{Robertson} B.,  {Goldreich} P.,  2012, \mn@doi [\apjl]
  {10.1088/2041-8205/750/2/L31}, \href
  {http://adsabs.harvard.edu/abs/2012ApJ...750L..31R} {750, L31}

\bibitem[\protect\citeauthoryear{{Sadavoy} et~al.,}{{Sadavoy}
  et~al.}{2010}]{Sadavoy_observed_CMF}
{Sadavoy} S.~I.,  et~al., 2010, \mn@doi [\apj] {10.1088/0004-637X/710/2/1247},
  \href {http://adsabs.harvard.edu/abs/2010ApJ...710.1247S} {710, 1247}

\bibitem[\protect\citeauthoryear{{Salpeter}}{{Salpeter}}{1955}]{Salpeter_slope}
{Salpeter} E.~E.,  1955, \mn@doi [\apj] {10.1086/145971}, \href
  {http://adsabs.harvard.edu/abs/1955ApJ...121..161S} {121, 161}

\bibitem[\protect\citeauthoryear{{Spaans} \& {Silk}}{{Spaans} \&
  {Silk}}{2000}]{Spaans00a}
{Spaans} M.,  {Silk} J.,  2000, \apj, 538, 115

\bibitem[\protect\citeauthoryear{{Stamatellos}, {Whitworth}  \&
  {Hubber}}{{Stamatellos} et~al.}{2012}]{Stamatellos_radfeedback}
{Stamatellos} D.,  {Whitworth} A.~P.,   {Hubber} D.~A.,  2012, \mn@doi [\mnras]
  {10.1111/j.1365-2966.2012.22038.x}, \href
  {http://adsabs.harvard.edu/abs/2012MNRAS.427.1182S} {427, 1182}

\bibitem[\protect\citeauthoryear{{Urban}, {Martel}  \& {Evans}}{{Urban}
  et~al.}{2010}]{Urban10a}
{Urban} A.,  {Martel} H.,   {Evans} N.~J.,  2010, \apj, 710, 1343

\bibitem[\protect\citeauthoryear{{Walch}, {Whitworth}  \& {Girichidis}}{{Walch}
  et~al.}{2012}]{Walch12a}
{Walch} S.,  {Whitworth} A.~P.,   {Girichidis} P.,  2012, \mnras, 419, 760

\bibitem[\protect\citeauthoryear{{Whitworth}, {Boffin}  \&
  {Francis}}{{Whitworth} et~al.}{1998}]{Whitworth98a}
{Whitworth} A.~P.,  {Boffin} H.~M.~J.,   {Francis} N.,  1998, \mnras, 299, 554

\bibitem[\protect\citeauthoryear{{van Dokkum} \& {Conroy}}{{van Dokkum} \&
  {Conroy}}{2010}]{Conroy_vanDokkum_ellipticals}
{van Dokkum} P.~G.,  {Conroy} C.,  2010, \mn@doi [\nat] {10.1038/nature09578},
  \href {http://adsabs.harvard.edu/abs/2010Natur.468..940V} {468, 940}

\makeatother
\end{thebibliography}

\end{document}